\documentstyle[preprint,aps]{revtex}
\input epsf.sty
\tighten

\begin{document}
\draft

\title{\Large \bf Monte Carlo Simulations of Short-time Critical
                  Dynamics with a Conserved Quantity }

\author{\bf B. Zheng$^{1,2}$ and H.J. Luo$^3$}

\address{$^1$FB Physik, Universit\"at Halle, D--06099 Halle, Germany}
\address{$^2$Institute of Theoretical Physics, Academia Sinica,
            100080 Beijing, P.R. China}
\address{$^3$FB Physik, Universit\"at Siegen, D--57068 Siegen, Germany}

\maketitle

\begin{abstract}
With Monte Carlo simulations, we investigate short-time
critical dynamics of the three-dimensional anti-ferromagnetic
Ising model with a globally conserved magnetization $m_s$
(not the order parameter). From the power law behavior
of the staggered magnetization (the order parameter),
its second moment and the auto-correlation,
we determine all static and dynamic critical exponents
as well as the critical temperature. The universality class
of $m_s=0$ is the same as that without a conserved quantity,
but the universality class of non-zero $m_s$ is different.

\end{abstract}

\pacs{PACS:  64.60.Ht, 02.70.Lq, 75.10.Hk}

\section{Introduction}

About ten years ago, in a pioneering work 
by Janssen, Schaub and Schmittmann, 
short-time universal
scaling behavior of non-equilibrium critical dynamics
starting from a high temperature state
was systematically explored with renormalization group methods
\cite {jan89}. 
In a few years, extension of the calculations
 to different critical dynamics 
was carried out by Oerding and Janssen in 
a series of works \cite {oer93,oer94,oer95}.
Some evidences for the short-time dynamic scaling
were also observed in Monte Carlo simulations
\cite {hus89,hum91,men94}.
Meanwhile, it was found that 
the power law decay of the magnetization starting
from a completely ordered state emerges already
in relatively early times, e.g., see \cite {sta92},
and it can be used
to estimate the dynamic exponent $z$
\cite {mue93,ito93}. In recent years
short-time critical dynamics
has been systematically investigated
with Monte Carlo simulations 
\cite {li94,sch95,gra95,zhe98,zhe99}.
Simulations have extended from regular classical spin models
\cite {zhe98,zha99,zhe99b,bra00,jen00}
to statistical systems with quenched disorder
\cite {luo98,luo99,yin00}, quantum spin systems
\cite {yin98}, lattice gauge theories \cite {oka98}, 
the hard-disk model \cite {jas99a,jas00} and dynamic
systems without detailed balance 
\cite {men96,men98,tom98,tom98a,bru99}. 
The references given here are only a part of recent ones
and not complete. A relatively complete list of
the relevant references before 1998 can be found
in a recent review, Ref. \cite {zhe98}.
All numerical and analytical results confirm that
there exists a rather general dynamic scaling form
in critical dynamic systems already in the short-time regime
of dynamic evolution.

Traditionally, it is believed that the short-time behavior
of dynamic systems depends essentially on microscopic
details. To understand short-time universal
behavior, one should distinguish two different time scales,
the macroscopic and microscopic time scales.
The short-time dynamic scaling form emerges only after
a time scale $t_{mic}$, which is sufficiently large
in microscopic sense, but still very small in macroscopic sense.
$t_{mic}$ is a time that the system needs to sweep
away the effect of microscopic details.
In Monte Carlo simulations, for example, if a sweep over
all spins on a lattice is considered to be
a microscopic time unit,
$t_{mic}$ is usually the order of 10 to 100 Monte Carlo time steps
\cite {zhe98}. For the simple Ising and Potts models
with only nearest neighbor interactions,
sometimes $t_{mic}$ is negligible small (e.g., one Monte Carlo
time step or two).
Compared with typical macroscopic time scales characterized
by $\tau^{-\nu z}$ or $L^z$ around the critical point, 
$t_{mic}$ is indeed very small.

The physical origin of the short-time dynamic scaling
is the divergent correlation time near a critical point.
The divergent correlation time induces a memory effect, and the memory effect
can be described by a scaling form. How does this scaling form
look like? As pointed out by Janssen, Schaub and Schmittmann
\cite {jan89}, one should introduce new critical exponents
to describe the dependence of the scaling behavior
on the {\it macroscopic initial conditions}. Actually, 
for arbitrary initial conditions even a characteristic function
is needed \cite {zhe96,che00}.
Furthermore, to describe the scaling behavior
of some special dynamic observables, 
we also have to introduce new exponents, for example,
the persistence exponent \cite {maj96a,sch97}.

The short-time dynamic scaling form is not only 
conceptually interesting but also practically important.
It provides
new techniques for the measurements of
both dynamic and static critical exponents as well as
the critical temperature, for a review, see Ref. \cite {zhe98}.
Since the measurements are carried out
in the short-time regime of dynamic
evolution, the dynamic approach does not suffer 
from critical slowing down.
\footnote {In the literature, sometimes it is stated
that the short-time dynamic approach can {\it eliminate}
critical slowing down. Rigorously speaking, this statement
is not correct. Critical slowing down always exists,
at least for local dynamics, and it is 
the physical origin of the dynamic scaling.
But in the short-time dynamic approach
we do not have the problem of generating
independent configurations.
Therefore, we do not suffer from critical slowing down.} 
Averaging is over both initial configurations
and the random forces. This is very different
from the time averaging in the measurements 
in equilibrium. Compared with those non-local methods 
which are invented to overcome the critical slowing down in
Monte Carlo simulations in equilibrium,
e.g., the cluster algorithms,
the dynamic approach does study the original local
dynamics and can be applied to disordered systems.

In fact, the dynamic scaling
in non-equilibrium critical dynamics
is not an exclusive phenomenon in the world.
In many other non-equilibrium dynamic processes,
universal or quasi-universal
scaling behavior has been also observed. An example is
 phase ordering dynamics \cite {bra94}. 
In this case, a real equilibrium state
and a macroscopic time scale
like $\tau^{-\nu z}$ do not exist. Therefore,
'short-time' is not addressed. Another example
is aging in complex systems such as glasses and spin glasses.
Aging is just a kind of scaling or quasi-scaling.
Dynamic scaling behavior around a spin glass transition
\cite {hus89,fis91,blu92,kis96,luo99}
is very similar to that around a standard critical point.
For example, the experimental measurements of the remanent
magnetization in spin glasses support not only the power law scaling behavior
but also the scaling relations between the 
exponents \cite {gra87,luo99}.

After such a slightly lengthy review,
we come to what we are going to do in this paper.
Even though renormalization group calculations
have been extended to different critical dynamics, up to now,
Monte Carlo simulations are limited in dynamics of model A
 \cite {hoh77}.
Actually, due to severe critical slowing down,
Monte Carlo simulations for long-time behavior of
critical dynamics beyond model A are also in 
 the preliminary stage. 
The short-time dynamic approach provides powerful methods
for numerical measurements of not only dynamic exponents, but
also static exponents as well as the critical temperature.
Simulations of dynamic systems beyond model A
in equilibrium
are usually difficult.
Therefore, it is important to investigate short-time
scaling behavior of these dynamic systems
with Monte Carlo methods.

Dynamics of model A is a kind of relaxational
dynamics {\it without} relevant conserved 
quantities \cite {hoh77}.
 If we consider only dynamic {\it relaxational} 
processes, dynamics with different relevant conserved quantities
is classified into model B, C and D.
In this paper, taking the three-dimensional 
anti-ferromagnetic Ising model for example,
with Monte Carlo methods we study 
dynamics {\it with a conserved quantity}, which is 
{\it not} an order parameter but coupling to the order parameter.
According to Ref. \cite {hoh77}, this is called model C.
In the next section, we introduce the model
and analyze the power law scaling behavior
in the short-time regime.
In section 3, we present the numerical results.
Finally come the concluding remarks.

\section{Short-time dynamic scaling}

\subsection{The model}

 We consider an antiferromagnetic
 Ising model on a three-dimensional cubic lattice.
 The Hamiltonian of the model is
\begin{equation}
-H/kT= - K \sum_{<ij>}  S_i S_j,
\label{e60}
\end{equation}
where $S_i$ is an Ising spin and the sum is over nearest neighbors.
For dynamics of model A and in equilibrium,
on a cubic lattice the antiferromagnetic
Ising model is equivalent to the (ferromagnetic)
Ising model.
Now, we keep {\it the magnetization
  as a constant in dynamic evolution.}
 Here the order parameter is {\it not }
 the magnetization but the staggered magnetization.
 We will denote the magnetization by $m_s$
 and the staggered magnetization by $M(t)$.
 For equilibrium states,
 it is expected that for $m_s=0$, the universality class
 is the same as that of the standard Ising model without
a conserved magnetization. For a non-zero $m_s$,
according to Ref. \cite {hoh77},
the critical exponent $\nu$ is different from that of $m_s=0.0$,
but it does not depend on the value of 
the non-zero $m_s$. For critical dynamics,
the dynamic exponent $z$ and the
 exponent $\theta$ for $m_s=0$ are the same as those
 of the standard
 Ising model, but for non-zero $m_s$ they
 are different. Again, $z$ and $\theta$ do not depend on
 the value of the non-zero $m_s$.
 These conclusions are drawn
 from renormalization group calculations
 based on the $\phi^4$ theory coupling to 
 a conserved current \cite {hoh77,oer93}.
 It is interesting to compared these results
 with Monte Carlo simulations.
 
 In this paper, we study only dynamic relaxation
 starting from {\it  disordered} states
 with a zero or small initial staggered magnetization $m_0$.
 We measure how the staggered magnetization 
 and its second moment as well as the auto-correlation
 evolve in the dynamic process.
The heat-bath algorithm is used in simulations.
 In order to keep $m_s$ as a constant,
 in a flip we simply exchange the values of {\it two} spins.
 Here we should point out that
 if our model is not the anti-ferromagnetic but
the ferromagnetic Ising model, the (conserved) 
 magnetization $m_s$ is the order
 parameter and then the dynamics for $m_s=0$ 
 belongs to model B \cite {hoh77}.
 
 In dynamics without any  relevant conserved quantities,
 i.e., dynamics of model A, updating schemes are irrelevant
 in the sense of universality.
 Different updating schemes lead to
 the same critical exponents, either in an equilibrium or
 a non-equilibrium state. However, for dynamics
 with relevant conserved quantities,
 it is different, at least for
 non-equilibrium short-time behavior.
 In a recent work by one of the authors
 \cite {zhe00}, for example, the Monte Carlo dynamics with a conserved
 order parameter for the two-dimensional Ising model
 is investigated. It is found that
 if in a flip we exchange the values of two {\it neighboring} spins,
 the dynamics is very slow. However, if we release
 the condition such that the order parameter is
 not locally conserved but only
 globally conserved (for example, in a flip
 we exchange the values of two randomly separated spins),
 the dynamic exponent $z$ is much smaller.
 For the dynamics studied in this paper, i.e.,
 with a conserved quantity
 which is not the order parameter,
 the situation is somewhat similar.
 If in a flip we exchange  the values of two neighboring spins,
 dynamic evolution is somewhat slow.
 Therefore, as a first approach, we decide to keep the magnetization $m_s$
 only globally conserved.
 In a flip we exchange the values of two randomly separated
 spins. Except for the updating schemes,
 all computational techniques adopted here
 are the same as those in the simulations of model A
 \cite {zhe98}.
 
\subsection{Short-time dynamic scaling}

For critical dynamic systems, 
traditionally it is believed that universal 
scaling behavior exists only in the long-time regime of 
dynamic evolution.
However, in recent years
 it is discovered that starting from  
{\it macroscopic} initial states, 
universal scaling behavior emerges already in the 
macroscopic short-time regime
of dynamic processes
after a microscopic time scale $t_{mic}$
\cite {jan89,hus89,hum91,li94,sch95,zhe98}.
A typical example is that 
a magnetic system initially in a high
temperature state with a {\it small} initial
order parameter $m_0$,
is suddenly quenched to the critical temperature $T_c$ 
or nearby
(without external magnetic field)
and then released to dynamic evolution of
model A \cite {hoh77}.
A generalized dynamic scaling form can be written down,
for example, for the $k$th moment of the order parameter,
\begin{equation}
M^{(k)}(t,\tau,L,m_0)=
b^{-k\beta/\nu}M^{(k)}(b^{-z}t,b^{1/\nu}\tau,b^{-1}L,b^{x_0}m_0).
\label{e15}
\end{equation}
Here $t$ is the time variable, $\tau$ is the reduced temperature,
$L$ is the lattice size. $\beta$ and $\nu$ are standard
static exponents and $z$ is the dynamic exponent.
Important is that a new independent exponent $x_0$
is introduced to describe the scaling behavior of 
the initial order parameter $m_0$.
If the scaling form above is valid,
 all relevant exponents
can be extracted from the short-time behavior of
relevant observables. 
 
 For the dynamic system with a conserved quantity
 discussed in the last subsection, we assume that a scaling form
 like Eq. (\ref {e15}) holds also.
 In this case, $\tau=(K-K_c)/K_c$ and $K_c$ depends on
 the conserved magnetization $m_s$, i.e., there is a critical line
 $K_c=K_c(m_s)$. In principle, there are two possibilities
 for the critical exponents. In the case of strong universality,
 the value of an exponent for non-zero $m_s$ can be different
 from that for zero $m_s$,
 but does not depend on the value of the non-zero $m_s$.
 In the case of weak universality, an exponent may vary
 its value continuously
 along a critical line. For the $\phi^4$ theory
 coupling to a conserved current in Ref. \cite {hoh77},
 it is the case of strong universality.

Neglecting the finite size effect
and noting that $m_0$ is small, we expand the right hand side
of Eq. (\ref {e15}) and take only the first non-zero (linear) term
under the condition of a small $t$.
At the initial stage of time evolution,
 the staggered magnetization around the critical temperature
behaves as,
\begin{equation}
M(t,\tau,m_0) \sim m_0 \, t^\theta F(t^{1/\nu z}\tau), 
 \qquad \theta=(x_0-\beta/\nu)/z .
\label{e10}
\end{equation} 
At the exact critical point, $\tau=0$,
$M(t)$ obeys a power law $\sim t^\theta$.
Numerical results and analytical calculations
have revealed that the exponent
$\theta$ is positive for almost all systems, i.e.,
the order parameter undergoes {\it an initial increase}.
This makes the short-time behavior very prominent.
The physical mechanism for this increase has not been
very clear. At least the mean-field effect or
symmetry breaking is not very relevant.
Slightly away from the critical point,
the power law behavior is modified by the scaling 
function $F(t^{1/\nu z}\tau)$.
This fact allows us to locate the critical temperature
and meanwhile to measure the critical exponent $\theta$.

Differentiation of Eq. (\ref {e10}) leads to
\begin{equation}
\partial _\tau M(t,\tau,m_0)|_{\tau=0}= m_0 \ t^{c_d}\ 
\partial _{\tau'}  F(\tau')|_{\tau'=0}, \qquad c_d=1 /\nu z+\theta.
\label{e80}
\end{equation}
From this power law behavior, one can 
determine the critical exponent $1/\nu z$.

Taking into account that the non-equilibrium
spatial correlation length ($\sim t^{1/z}$) 
is small at the initial stage
of time evolution, from Eq. (\ref {e15}) it  
can be derived that
the second moment  at $T_c$
subjects to a finite size scaling
\begin{equation}
M^{(2)}(t,L) \sim L^{-d}\; t^y\;, \qquad
y=(d-2\beta/\nu )/z.
\label{e40}
\end{equation}
For simplicity, here we have set $m_0=0$.
Another interesting dynamic observable is the auto-correlation 
\begin{equation}
A(t) \equiv \frac{1}{L^d}\, \langle\sum_i S_i(0) S_i(t)\rangle \ .
\label{e20}
\end{equation}
At the critical temperature $T_c$, $A(t)$
decays by a  power law \cite {jan92}
\begin{equation}
A(t) \sim t^{-\lambda}, \quad \lambda=\frac{d}{z}-\theta.
\label{e30}
\end{equation}
It is interesting that even though we have set
$m_0=0$, $\theta$ (i.e., $x_0$) still enters
the auto-correlation. This is because $x_0$ is actually
the scaling dimension of the {\it local} order parameter.

From Eqs. (\ref {e40}) and (\ref {e30}), we are able to estimate the
static exponent $\beta/\nu$ and the dynamic exponent $z$.
Then we complete the measurements of all the exponents
and the critical temperature.
For details of the above scaling analysis and
a more systematic extension,
readers are referred to Refs.~\cite {jan89,jan92}
and the recent review article Ref.~\cite {zhe98}.

\section{Monte Carlo simulations}

\subsection{$m_s=0$}

For the anti-ferromagnetic Ising model,
in equilibrium the averaged magnetization
is zero. If we notice that the order parameter is
the staggered magnetization, the thermodynamic fluctuation
of the magnetization
goes to zero in the thermodynamic limit
(infinite volume).
In non-equilibrium dynamic processes
without any conserved quantities, if one starts from 
initial states with a zero magnetization $m_s$,
the magnetization will remain zero.
This is the case for usual macroscopic initial states,
e.g., the high temperature and low temperature states.
The fluctuation of the magnetization in the dynamic processes
is also zero in thermodynamic limit.
Therefore, one expects that the equilibrium state
of the dynamic process with a conserved magnetization $m_s=0$
is in a same universality class of the model without 
a conserved quantity,
and so is the dynamic universality class as well.
The critical temperature is also not changed.

We perform simulations for $m_s=0$ in order to confirm 
the above expectation and to demonstrate our 
computational techniques. The heat-bath algorithm
is adopted in simulations since it is faster
than the Metropolis algorithm in the short-time regime.
Lattice sizes are $L=64$ and $128$.
Within our maximum updating times, no finite size effects are observed.
We remind the readers here that
it is an advantage of the short-time dynamic approach
that the finite size effects are easily controlled
due to the small non-equilibrium
spatial correlation length ($\sim t^{1/z}$).

The critical temperature has been measured rather accurately
in simulations in equilibrium (for the system
without a conserved quantity). However,
we also present data to determine it from
short-time dynamics. With the short-time dynamic approach, in principle,
any observables which are sensitive to the temperature
around the critical regime can be used for the determination
of the critical temperature.

\begin{figure}[t]\centering 
\epsfysize=10.cm 
\epsfclipoff 
\fboxsep=0pt
\setlength{\unitlength}{1cm} 
\begin{picture}(13.6,12.)(0,0)
\put(-1.,0){{\epsffile{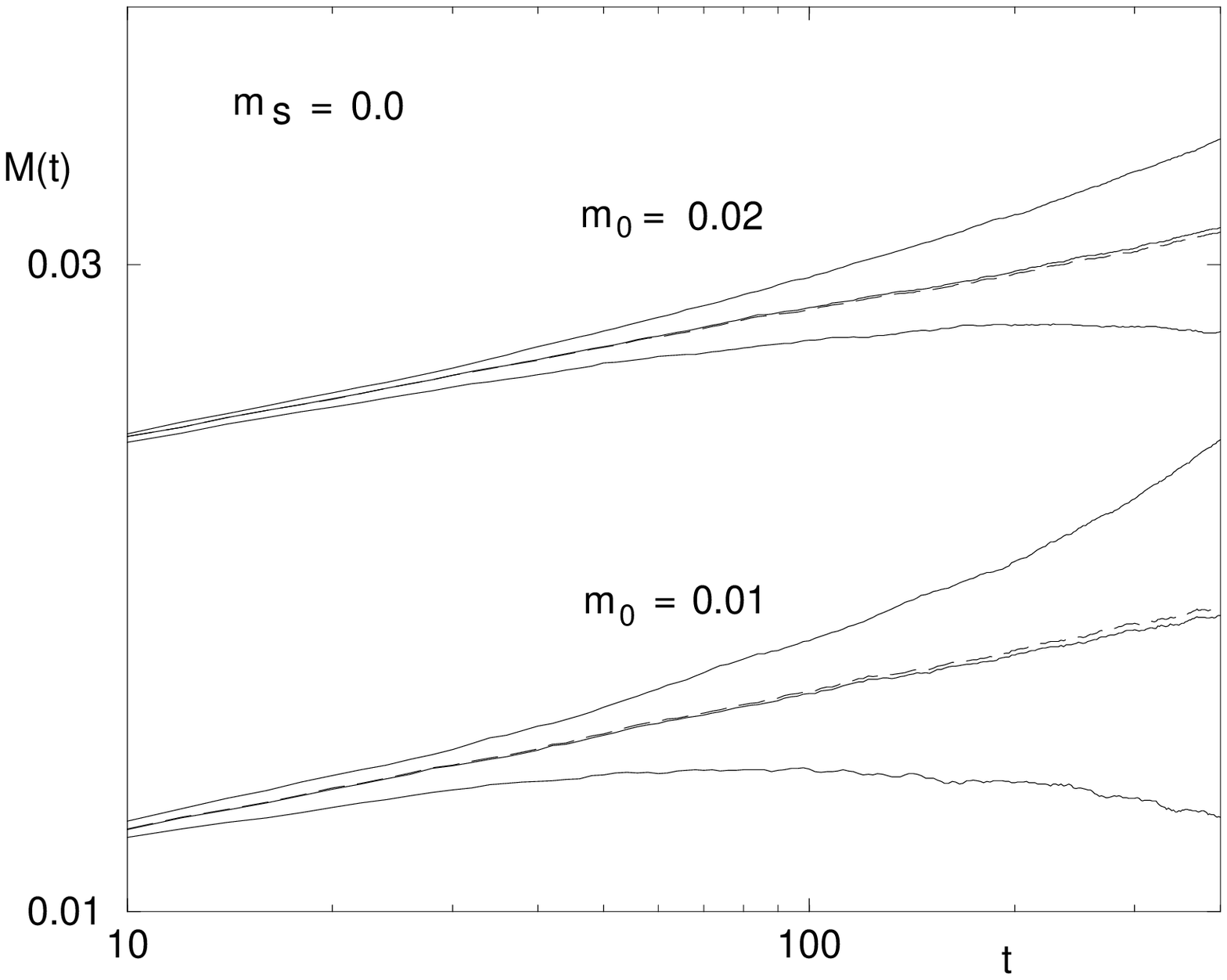}}} 
\end{picture} 
\caption{Time evolution of the staggered magnetization
displayed in log-log scale. For solid lines, $K$
 is $0.22065$, $0.22165$ and $0.22265$ 
 (from below) for $m_0=0.01$, and 
$K$ is $0.22115$, $0.22165$ and $0.22215$ for $m_0=0.02$.
 Dashed lines correspond to $K_c=0.22169(9)$ and
 $0.22163(5)$ for $m_0=0.01$ and $m_0=0.02$.
} 
\label{f1}
\end{figure}

In Fig. \ref {f1}, time evolution of the staggered magnetization
$M(t)$ is displayed with solid lines in log-log scale.
The lattice size is $L=128$, and $2500$ samples
of initial configurations have been used for averaging.
In order to see the possible 'finite $m_0$ effects',
we have performed the simulations with  
$m_0=0.01$ and $0.02$. The coupling constant $K$
 is $0.22065$, $0.22165$ and $0.22265$ 
 (from below) for $m_0=0.01$, and 
 $0.22115$, $0.22165$ and $0.22215$ for $m_0=0.02$.
In principle, the closer $K$'s are used,
 the more accurate $K_c$ and the critical
 exponent $1/\nu z$ can be obtained. However,
 we suffer from large statistical fluctuation if
 the $K$'s are too close to each other. We use different $K$'s
 for $m_0=0.01$ and $m_0=0.02$ just to study 
 the possible systematic errors.

 To make use of Eq. (\ref {e10}) to locate the critical point,
 we first interpolate $M(t)$ quadratically to 
 any $K$ around the three simulated ones,
 then search for a $K$ which gives the best power law behavior
 for the curve. This $K$ is $K_c$.
 The two dashed lines in Fig. \ref {f1}
 are the curves with the best power law behavior
 for $m_0=0.01$ and $m_0=0.02$. The corresponding $K_c$'s are
 $K_c=0.22169(9)$ and $0.22163(5)$
 for $m_0=0.01$ and $m_0=0.02$ respectively. These two values are
 consistent each other and
 in agreement with that of the Ising model measured in equilibrium
  (without any conserved quantity), which is reported to
 be around $0.22165$ \cite {fer91,blo95}.
 We perform our measurements in a time interval
 $[t_{mic},400]$. The results are rather stable
 when we take $t_{mic}$ bigger than $10$.

With $K_c$ in hand, we measure the critical exponent 
$\theta=0.108(5)$ and $0.100(5)$ for
$m_0=0.01$ and $m_0=0.02$. They are consistent
with $\theta=0.104(3)$ in Ref. \cite {gra95}
and $\theta=0.108(2)$ in Ref. \cite {jas99}
for the Ising model without a conserved quantity.
Rigorously speaking, the exponent $\theta$
is defined in the limit of $m_0=0$. We should extrapolate
$\theta$ to $m_0=0$. However, since here there are some
statistical errors, and especially the errors induced by
the errors of the critical point $K_c$'s,
it is not meaningful to do so. To reduce the errors
induced by $K_c$, we could take $K_c=0.22165$ as input.
Then we obtain $\theta=0.106(6)$ by extrapolating
the results to $m_0=0$. In this paper, 
we simply consider $\theta=0.108(5)$ measured from 
$m_0=0.01$ as our final value of $\theta$.

\begin{figure}[t]\centering 
\epsfysize=10.cm 
\epsfclipoff 
\fboxsep=0pt
\setlength{\unitlength}{1cm} 
\begin{picture}(13.6,12.)(0,0)
\put(-1.,0){{\epsffile{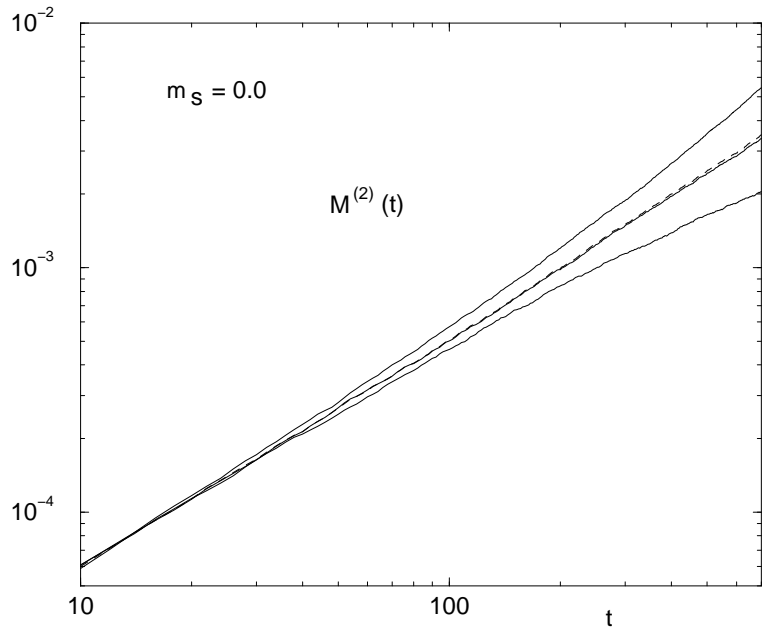}}} 
\end{picture} 
\caption{The second moment in log-log scale. For solid lines, $K$
 is $0.22065$, $0.22165$ and $0.22265$ 
 (from below).
 The dashed line corresponds to $K_c=0.22164(22)$.
} 
\label{f2}
\end{figure}

In Fig. \ref {f2}, time evolution of the second moment
of the staggered magnetization
is displayed with solid lines in log-log scale.
Here the initial staggered magnetization $m_0$
has been set to zero.
From below, $K=0.22065$, $0.22165$ and $0.22265$.
The lattice size is $L=64$ and the number of samples for averaging
is $3200$. Extra simulations for $L=128$ show that
the finite size effect for $L=64$ is already
negligible small. From the figure, we see that 
the second moment is apparently less sensitive
to $K$, compared with the staggered magnetization itself.
However, we still can locate the critical $K_c$ 
from the data. Measuring in a time interval 
$[t_{mic}=20,700]$, $K_c=0.22164(22)$.
The curve corresponding to $K_c$ is shown by a dashed line
in the figure.
The error of $K_c$ here
is bigger than that from the staggered magnetization.
To reduce the error, simply increasing the samples
for averaging is not enough, one must 
have longer maximum updating times and therefore larger
lattices. These need much CPU times since the
second moment is not self-averaging.
To estimate the exponent $y=(d-2\beta/\nu )/z$,
for simplicity, we take $K_c=0.22165$ as input.
The result is $y=0.965(11)$.

\begin{figure}[t]\centering 
\epsfysize=10.cm 
\epsfclipoff 
\fboxsep=0pt
\setlength{\unitlength}{1cm} 
\begin{picture}(13.6,12.)(0,0)
\put(-1.,0){{\epsffile{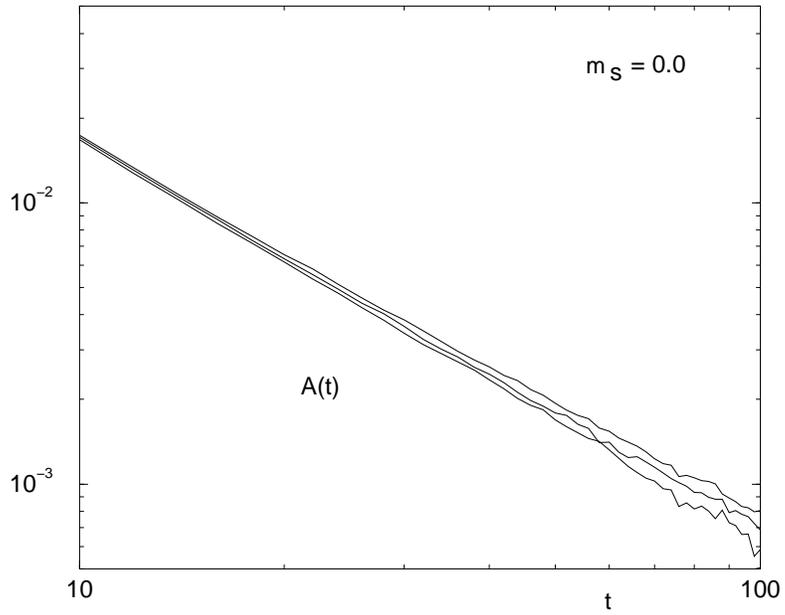}}} 
\end{picture} 
\caption{The auto-correlation in log-log scale. $K$
 is $0.22065$, $0.22165$ and $0.22265$ 
 (from below).
} 
\label{f3}
\end{figure}

In the simulations for Fig. \ref {f2}, we have also measured
the auto-correlation function $A(t)$.
The results are shown in Fig. \ref {f3}.
From below, $K=0.22065$, $0.22165$ and $0.22265$.
It is impossible to locate the critical point
from the auto-correlation. We measure the exponent
$\lambda=1.36(3)$ at $K_c=0.22165$ in a time
interval $[20,100]$. When $t$ is bigger than $100$,
the fluctuation becomes large.

\begin{figure}[t]\centering 
\epsfysize=10.cm 
\epsfclipoff 
\fboxsep=0pt
\setlength{\unitlength}{1cm} 
\begin{picture}(13.6,12.)(0,0)
\put(-1.,0){{\epsffile{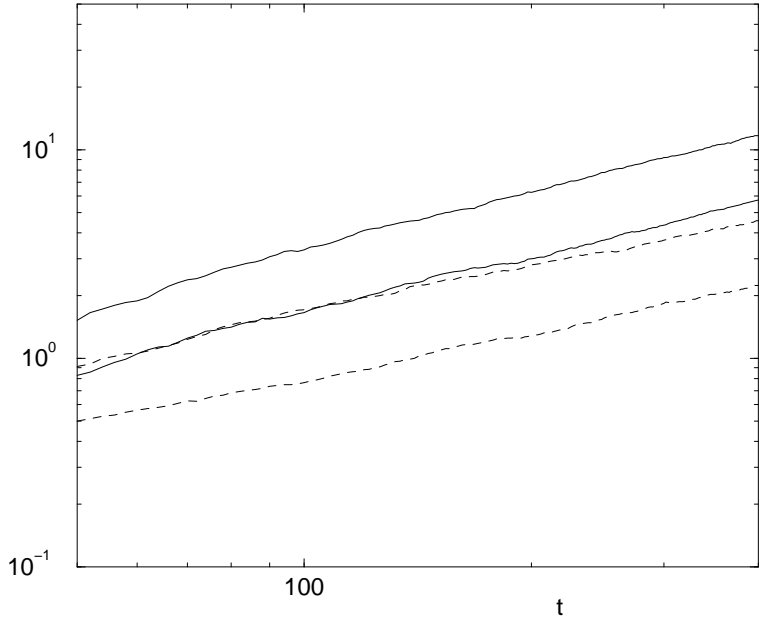}}} 
\end{picture} 
\caption{The derivative $\partial _\tau M(t,\tau)|_{\tau=0}$
in log-log scale. Solid lines are for $m_s=0.0$ and 
 from below, $m_0=0.01$ and $0.02$. Dashed lines are for 
 $m_s=0.2$ and $0.4$ (from above) with $m_0=0.01$.
} 
\label{f4}
\end{figure}

In order to determine the critical exponent $\nu$,
we need 
$\partial _\tau M(t,\tau)|_{\tau=0}$. 
From the data for Fig. \ref {f1}, we can calculate this 
derivative approximately. The results are shown
in Fig. \ref {f4} with solid lines in log-log scale.
From below, the initial staggered magnetization
is $m_0=0.01$ and $0.02$. It is known that the power law
behavior of the derivative of $M(t)$ in shorter times 
is less clean than that of $M(t)$ itself \cite {zhe98}.
Therefore, we measure the slopes in a time interval
$[70,400]$ and obtain
$c_d=1/\nu z + \theta=0.867(10)$ and $0.901(22)$
for $m_0=0.01$ and $0.02$ respectively.
The value from $m_0=0.01$ is more reliable.

In Table \ref {t1}, we list all the exponents
we have measured. Taking the exponent $\theta$ as input,
from $\lambda$ we estimate the dynamic exponent $z$.
Then from $y$ and $c_d$ we calculate the exponents
$2 \beta/\nu$ and $\nu$. For comparison,
available results for dynamics without a conserved quantity
\cite {jas99},
i.e., the so-called model A dynamics, are also given,
where the critical point $K_c$ and the exponent $\nu$
are measured from another dynamic process starting from
a completely ordered state. Within statistical errors,
all the exponents for dynamics with and without
a conserved quantity agree well each other.
The static exponents and $K_c$ are also consistent
with those of the standard Ising model 
measured from simulations in equilibrium
with the cluster algorithms \cite {blo95}.
Here we should make a comment.
Even for dynamics of model A, the static exponents and $K_c$
measured from the short-time dynamics are still not 
as accurate as those obtained with the cluster algorithms.
One reason is that the best effort for the short-time
dynamic approach has not been made. Another reason is that
in the short-time dynamic approach we have to measure
both the static and dynamic exponents together.
It makes the work more difficult than measuring
only the static exponents. However, it is clear
if we are interested in both static and dynamic properties,
the short-time dynamic approach is very efficient.
The cluster algorithms are non-local and change already
the dynamic universality class. 
On the other hand, the cluster algorithms 
can not straightforwardly apply to any systems,
e.g., the systems with quenched randomness and
lattice gauge theories.

\subsection{$m_s\neq 0$}

Encouraged by the success for $m_s=0.0$, we proceed to 
$m_s\neq 0$. The results in the last subsection show
that the lattices sizes $L=64$ and $128$ are sufficient
for our maximum updating times. The finite $m_0$
effect for $m_0=0.01$ is already invisible within
the statistical errors. Therefore, in this subsection
we will not systematically study the finite size
and finite $m_0$ effects. However we have also performed
some extra calculations to confirm the results presented.
To locate the critical point $K_c$ and estimate
the exponent $\nu$, we need data for several $K$'s 
in the neighborhood of the critical point.
The results in the last subsection also show that
the difference of these $K$'s should be 
around half a percent of $K_c$ or less.

\begin{figure}[t]\centering 
\epsfysize=10.cm 
\epsfclipoff 
\fboxsep=0pt
\setlength{\unitlength}{1cm} 
\begin{picture}(13.6,12.)(0,0)
\put(-1.,0){{\epsffile{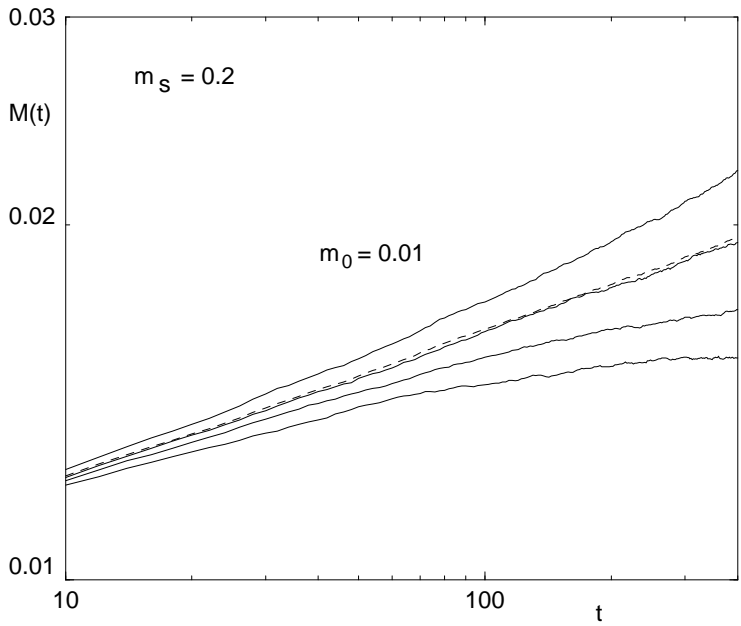}}} 
\end{picture} 
\caption{Time evolution of the staggered magnetization
for $m_s=0.2$ in log-log scale . For solid lines, $K$
 is $0.2405$, $0.2410$, $0.2415$ and $0.2420$ 
 (from below). 
 The dashed line corresponds to $K_c=0.24153(12)$.
} 
\label{f5}
\end{figure}

\begin{figure}[t]\centering 
\epsfysize=10.cm 
\epsfclipoff 
\fboxsep=0pt
\setlength{\unitlength}{1cm} 
\begin{picture}(13.6,12.)(0,0)
\put(-1.,0){{\epsffile{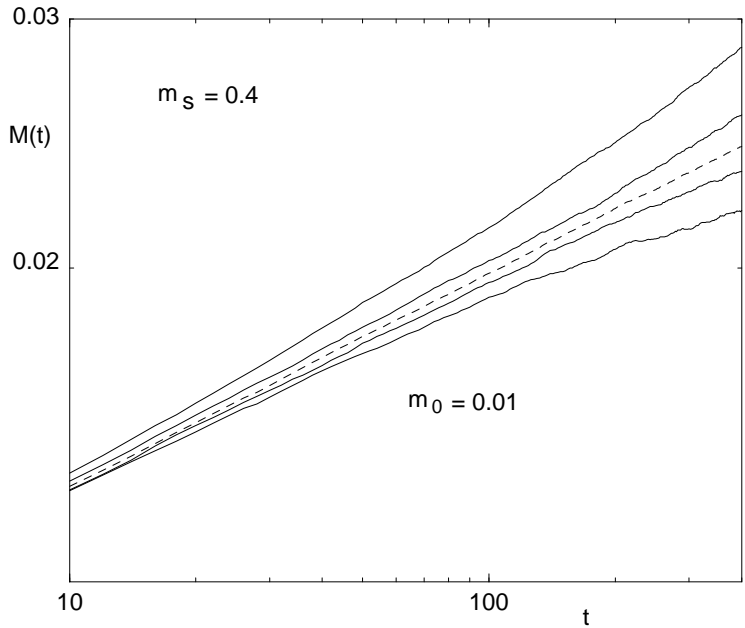}}} 
\end{picture} 
\caption{Time evolution of the staggered magnetization
for $m_s=0.4$ in log-log scale. For solid lines, $K$
 is $0.3310$, $0.3320$, $0.333$ and $0.3340$ 
 (from below). 
 The dashed line corresponds to $K_c=0.33230(20)$
} 
\label{f6}
\end{figure}

  We perform simulations for $m_s=0.2$ and $0.4$.
  For $m_s=0.2$ we obtain data for
  $K=0.2405$, $0.2410$, $0.2415$ and $0.2420$,
  which are shown in Fig. \ref {f5} with solid
  lines (from below) in log-log scale.
  For $m_s=0.4$ we obtain data for
  $K=0.3310$, $0.3320$, $0.3330$ and $0.3340$,
  which are shown in Fig. \ref {f6} with solid
  lines (from below) in log-log scale.
   The lattice size is $L=128$
  and  $2500$ samples of the initial configurations
  are used for averaging. 
  
  Following the procedure
  in the last subsection, searching for a curve
  with the best power law behavior in a time interval
  $[t_{mic}, 400]$, one can determine $K_c$ and
  the critical exponent $\theta$.
  Now we have data for four $K$'s and therefore
  have the choices to interpolate the staggered magnetization
  with three or four $K$'s. For $K_c$,
  both yield the same values within errors.
  With four $K$'s, we obtain 
  $K_c=0.24153(12)$ for $m_s=0.2$ and 
  $K_c=0.33230(20)$ for $m_s=0.4$ with
  $t_{mic}=20$. The corresponding curves
  are shown in Figs. \ref {f5} and \ref {f6} with dashed lines.
  For $m_s=0.4$, $K_c$ is very stable
  for different choices of $t_{mic}$.
  For $m_s=0.2$, $K_c$ increases slightly
  when we take bigger $t_{mic}$. The reason
 is that our maximum updating times are not long enough.
 After carefully analyzing the data and looking at 
 the curves in the figure, we convince ourself
 that $K_c=0.24153(12)$ is the most reasonable.
  
  In Fig. \ref {f4}, $\partial _\tau M(t,\tau)|_{\tau=0}$ 
  is displayed with dashed lines for $m_s=0.2$ and
  $0.4$ (from above). The results are obtained with four $K$'s.
  For $m_s=0.4$, the resulting exponent
  $\nu$ is bad with three $K$'s. The reason is clear.
  $K_c$ is between $K=0.3320$ and $0.3330$.
  If we choose only three $K$'s, we have some systematic
  errors in interpolation.

\begin{figure}[t]\centering 
\epsfysize=10.cm 
\epsfclipoff 
\fboxsep=0pt
\setlength{\unitlength}{1cm} 
\begin{picture}(13.6,12.)(0,0)
\put(-1.,0){{\epsffile{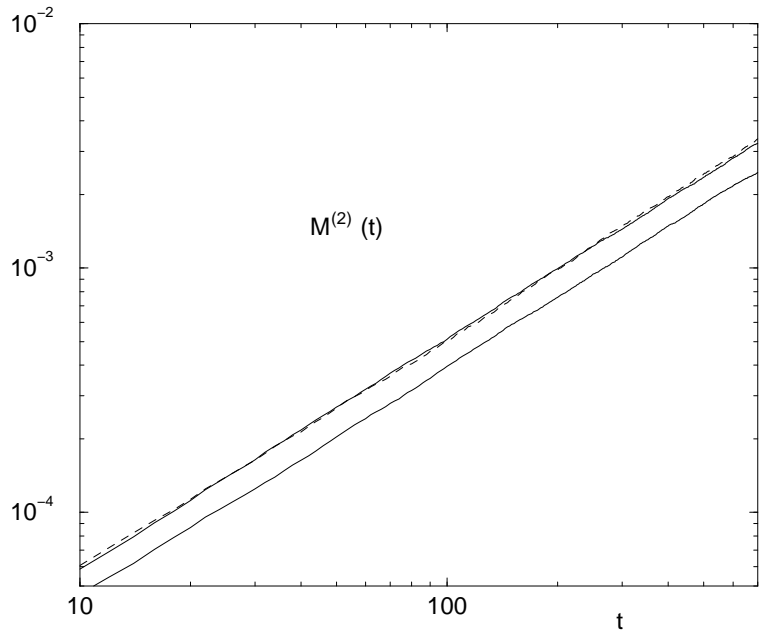}}} 
\end{picture} 
\caption{The second moment at $K_c$
in log-log scale. For solid lines, 
 $m_s=0.2$ and $0.4$ 
 (from above).
 The dashed line is for $m_s=0.0$.
} 
\label{f7}
\end{figure}

\begin{figure}[t]\centering 
\epsfysize=10.cm 
\epsfclipoff 
\fboxsep=0pt
\setlength{\unitlength}{1cm} 
\begin{picture}(13.6,12.)(0,0)
\put(-1.,0){{\epsffile{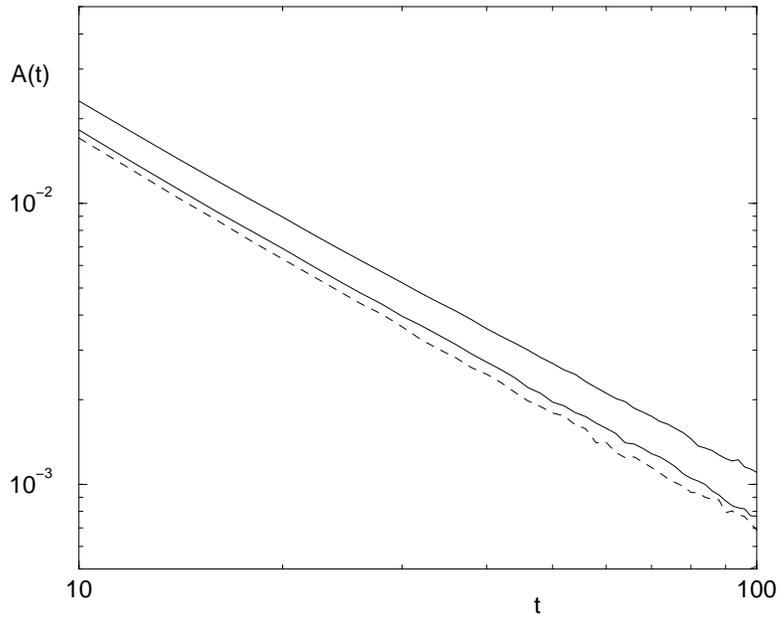}}} 
\end{picture} 
\caption{The auto-correlation at $K_c$
in log-log scale. For solid lines, 
 $m_s=0.2$ and $0.4$ 
 (from below).
 The dashed line is for $m_s=0.0$.
} 
\label{f8}
\end{figure}

  With $K_c$ in hand, we perform simulations for $m_0=0.0$
  and measure the second moment and the auto-correlation.
  A lattice size $L=64$ is used and the number of samples of initial 
  configuration for averaging is $7500$.
  In Fig. \ref {f7}, the second moment is displayed
  in log-log scale with solid lines for 
  $m_s=0.2$ and $m_s=0.4$ (from above).
  For comparison, the second moment for $m_s=0.0$
  is also plotted with a dashed line in the figure.
  Apparently, all three curves are parallel each other.
  The exponent $y$ is independent of $m_s$.
  In Fig. \ref {f8}, the auto-correlation is displayed
  in log-log scale with solid lines
  for $m_s=0.2$ and $m_s=0.4$ (from below) and
  with a dashes line for $m_s=0.0$. 
  In all the relevant figures for non-zero $m_s$,
  nice power law behavior is observed.

  In Table \ref {t1}, all the exponents are summarized.
  Within the statistical errors,
  the static exponent $2 \beta /\nu$
  is independent of $m_s$. The values also
  coincide well with those obtained from dynamics of model A and
  measured in equilibrium for the standard Ising model.
   The static exponent
  $\nu$ for $m_s=0.2$ and $0.4$ are
  $0.82(8)$ and $0.81(3)$ respectively. Even though the first
  value carries a relatively big error, they are clearly
  different from $\nu=0.64(2)$ for $m_s=0.0$
  and those measured from dynamics of model A and in equilibrium.
  However, $\nu$ is independent of the value
  of the non-zero $m_s$. These results
  are consistent with the theoretical calculations
  based on a $\phi^4$ theory coupling to a conserved
  current \cite {hoh77}.
  With our accuracy, we can not detect any dependence
  of the dynamic exponent $z$ on $m_s$.
  This differs from the $\phi^4$ theory in Ref. \cite {hoh77}.
  The exponent $\theta$ is $0.129(5)$ and $0.148(4)$
  for $m_s=0.2$ and $0.4$ respectively . These two values
  are obviously different
  from $\theta=0.108(5)$ for $m_s=0.0$,
and show the nontrivial dynamic behavior for nonzero
$m_s$. However,
  with our data we should conclude that $\theta$ depends on the value
  of the non-zero $m_s$, even though there may be still 
  some uncontrolled systematic errors.
  This point is different from
  the $\phi^4$ theory.

\section{Concluding remarks}

We have reported our Monte Carlo simulations of short-time
critical dynamics for the three-dimensional 
anti-ferromagnetic Ising model with a globally conserved
magnetization (not an order parameter). The power law
scaling behavior has been observed for the staggered magnetization
(the order parameter), its second moment and the
auto-correlation. All dynamic and static critical exponents
are determined. For a conserved magnetization $m_s=0.0$,
all the exponents are the same as those of model A.
For a non-zero $m_s$, the static exponents
behave qualitatively the same as those of
the $\phi^4$ theory coupling to a conserved
current \cite {hoh77}. However, the dynamic exponents
are somewhat different.
The reason might be that the magnetization is not locally
but only globally conserved. Therefore, it is interesting
to study the dynamics, with a locally conserved order parameter
or non-order parameter: 
in a flip, we exchange two spins in a local regime,
the size of which must be much smaller than
the lattice size.

{\bf Acknowledgements}:
B.Z. thanks deeply Oerding for suggesting the topic
and inspiring discussions. 
This work is supported in part by DFG; TR 300/3-1 and Schu 95/9-2.


\begin{table}[h]\centering 
\begin{tabular}{l|lllll|lll}
 $m_s$ & $K_c$ & $\theta$ & $\lambda$ &  $y$ & $c_d$  & $z$   
     & $ 2\beta/\nu$  & $\nu$ \\ 
 \hline 
 0.0 & 0.22169(9) &  0.108(5) & 1.36(3)  &  0.965(11) & 0.867(10)
     &  2.04(4)   &  1.03(4)  &  0.64(2) \\ 
 \hline 
 0.2 & 0.24153(12) &  0.129(5) & 1.34(2)  &  0.954(4) & 0.725(70)
     &  2.04(3)   &  1.05(3)  &  0.82(8) \\ 
 \hline 
 0.4 & 0.33230(20) &  0.148(4) & 1.31(1)  &  0.945(8) & 0.748(20)
     &  2.06(2)   &  1.05(3)  &  0.81(3) \\ 
\hline
 model A & 0.22170(4) &  0.108(2) & 1.36(2)  &  0.970(11) & 
     &  2.04(2)   &  1.034(4)  &  0.633(3) \\  
\hline
 Ising & 0.221655(1) &  &   &   & 
     &    &  1.037(3)  &  0.630(1)   
\end{tabular} 
\caption{Critical exponents measured for different conserved magnetization
$m_s$'s.
The values for dynamics of model A are taken from Ref.
\protect \cite {jas99}. Under the item 'Ising'
are the values for the standard Ising model in equilibrium
\protect \cite {blo95}.  } 
\label{t1} 
\end{table}

\end{document}